\documentclass[twocolumn,showpacs,preprintnumbers,amsmath,amssymb]{revtex4}
\usepackage{graphicx}
\usepackage{dcolumn}
\usepackage{amsmath}
\def\apj #1 #2 #3 {#1, ApJ, {\bf #2}, #3}
\def\apjl #1 #2 #3 {#1, ApJ, {\bf #2}, L#3}
\def\apjs #1 #2 #3 {#1, ApJS, {\bf #2}, #3}
\def\aap  #1 #2 #3 {#1, A\&A, {\bf #2}, #3}
\def\mnras #1 #2 #3 {#1, MNRAS, {\bf #2}, #3}
\def\pra #1 #2 #3 {#1, Phys.~Rev.~A., {\bf #2}, #3}
\def\prb #1 #2 #3 {#1, Phys.~Rev.~B., {\bf #2}, #3}
\def\prc #1 #2 #3 {#1, Phys.~Rev.~C., {\bf #2}, #3}
\def\prd #1 #2 #3 {#1, Phys.~Rev.~D., {\bf #2}, #3}
\def\pre #1 #2 #3 {#1, Phys.~Rev.~E., {\bf #2}, #3}
\def\prl #1 #2 #3 {#1, Phys.~Rev.~Lett., {\bf #2}, #3}
\def\plb #1 #2 #3 {#1, Phys.~Lett.~B., {\bf #2}, #3}
\def\science #1 #2 #3 {#1, Science., {\bf #2}, #3}
\def\nature #1 #2 #3 {#1, Nature., {\bf #2}, #3}
\def\nphysa #1 #2 #3 {#1, Nucl.~Phys.~A., {\bf #2}, #3}
\def\nphysb #1 #2 #3 {#1, Nucl.~Phys.~B., {\bf #2}, #3}
\def\nphysbs #1 #2 #3 {#1, Nucl.~Phys.~B.~Suppl., {\bf #2}, #3}

\def\h#1{\hbox{${}^{#1}$H}}

\def\h502{\hbox{$ h^{2}_{50}$}}

\def\fun#1#2{\lower3.6pt\vbox{\baselineskip0pt\lineskip.9pt
  \ialign{$\mathsurround=0pt#1\hfil##\hfil$\crcr#2\crcr\sim\crcr}}}
\begin{document}
%
\title{Disappearing Dark Matter in Brane World Cosmology: 
 New Limits on Noncompact Extra Dimensions }
\author{K. Ichiki$^{1,2}$, P. M. Garnavich$^{3}$,
 T. Kajino$^{1}$, G. J. Mathews$^{3}$, and M. Yahiro$^{4}$}
\affiliation{
$^1$National Astronomical Observatory, 2-21-1, Osawa, Mitaka, Tokyo
181-8588,
Japan}
\affiliation{
$^2$University of Tokyo, Department of Astronomy, 7-3-1
Hongo, Bunkyo-ku, Tokyo 113-0033, Japan }
\affiliation{$^3$Center for Astrophysics,
 University of Notre Dame, Notre Dame, IN 46556}
\affiliation{
$^4$Department of Physics and Earth Sciences, University of the Ryukyus,
Nishihara-chou, Okinawa 903-0213, Japan }
\date{\today}
\begin{abstract}
 We explore cosmological implications of dark matter as massive
 particles trapped on a brane embedded in a Randall-Sundrum noncompact
 higher dimension $AdS_5$ space. It is an unavoidable consequence of
 this cosmology that massive particles are metastable and can disappear
 into the bulk dimension. Here,  we show that a massive dark matter
 particle (e.g. the lightest supersymmetric particle) is likely to have
 the shortest lifetime  for disappearing into the bulk. We examine
 cosmological constraints on this new paradigm and show that
 disappearing dark matter is consistent (at the  95\% confidence level)
 with all cosmological constraints, i.e. present observations of Type Ia
 supernovae at the highest redshift, trends in the mass-to-light ratios
 of galaxy   clusters with redshift, the fraction of X-ray emitting gas
 in rich clusters, and the spectrum of power fluctuations in the cosmic
 microwave background.  A best $2 \sigma$ concordance region is
 identified corresponding to a mean lifetime for dark matter
 disappearance of $15 \le \Gamma^{-1} \le 80$ Gyr. The implication of
 these results for brane-world physics is discussed.
\end{abstract}

\maketitle
%
%
%

\section{INTRODUCTION}
There is currently considerable interest in the possibility that our
universe could be a submanifold embedded in a higher-dimensional
spacetime. This brane-world paradigm is motivated by the D-brane
solution found in ten-dimensional superstring theory. Technically, in
type IIB superstrings, an $AdS_5 \times S^5$ geometry is formed near the
stacked D3-branes \cite{M1,GKP1,W1,Poly1}.  In simple terms this means
that a model can be proposed \cite{Randall} whereby our universe is
represented as a thin three-brane embedded in an infinite
five-dimensional bulk anti-de Sitter space ($AdS_5$). In such
Randall-Sundrum (RSII) models, physical particles are trapped on a
three-dimensional brane via curvature in the bulk dimension. Gravitons
can reside as fluctuations in the background gravitational field living
in both the brane and bulk dimension. This representation of large extra
dimensions is an alternative to the standard Kaluza-Klein  (KK)
compactification. 

Although massive particles can indeed be trapped on
the brane, they are also, however,  expected to be metastable
\cite{dubovsky}. That is, for both scalar and fermion fields, the
quasi-normal modes are metastable states that can decay into continuum
KK modes in the higher dimension. From the viewpoint of an observer on
the three-brane, massive particles will appear to propagate for some
time and then literally disappear into the bulk fifth dimension.

In the RSII model, curvature in the bulk dimension is introduced as a
means to suppress the interaction of massless particles with the
continuum of KK states in the bulk dimension. However, introducing a
mass term into the higher-dimensional action leads to nonzero coupling
to that KK continuum. The mathematical realization of this decay is
simply  that the eigenvalues for the mass modes of the field theory are
complex.

The simplest  model to illustrate this is the case of a free scalar
field to which a bulk mass term $\mu$ has been added \cite{dubovsky}. In
this case, the imposition of radiation (outgoing-wave) boundary
conditions on the solution to the five dimensional  Klein-Gordon
equation leads to complex eigenvalues of the form, 
\begin{equation}
m =m_0 - i \Gamma ~~,
\end{equation}
where, quasi-discrete four-dimensional masses are given by
\begin{equation}
m_0^2 = \mu^2/2~~,
\end{equation}
with $\mu$ being the bulk mass term in the $AdS_5$ field equation.
The width $\Gamma$ is given by
\begin{equation}
{\Gamma} = ({\pi}/{16}) ({m_0^3}/{L^2})~~,
\label{gammas}
\end{equation}
where $L$ is the metric curvature parameter of the bulk dimension.
 That is, we write the five-dimensional metric,
\begin{equation}
ds^2 = \exp^{-2 \vert z \vert L} \eta_{\mu \nu} dx^\mu dx^\nu + dz^2~~,
\end{equation}
where $z$ is the bulk dimension and the bulk curvature parameter is,
\begin{equation}
L  = \sqrt{-\Lambda_5/6} ~~,
\label{ldef}
\end{equation}
where, $\Lambda_5$ is the negative bulk cosmological constant. A
construction of the propagator for particles on the brane then has a
pole at complex $p^2$ which corresponds to an unstable particle with
mass $m_0$ and width $\Gamma$. Thus, the comoving density of massive
scalar particles can be expected to decay over time with a rate, $(\rho
a^3)\exp{[- \Gamma t ]}$, where $a$ is the scale factor.

It is well known \cite{Bajc} that fermion fields cannot be localized on
a brane with positive tension by gravitational interactions only.  One
must invoke a localization mechanism.  A simple example \cite{dubovsky}
is to form a domain wall by introducing a scalar field $\chi$ with two
degenerate vacua $\chi = \pm v$ separated by a domain wall at the
brane. A fermion field is then introduced with a Yukawa coupling to the
scalar field, $g \chi \bar \psi \psi$, which confines fermions to the
brane.  Similar to the treatment of scalar particles, solving the Dirac
equation for fermions with a bulk mass term $\mu$, leads to complex mass
eigenvalues. In the limit that the bulk mass is much less than the
curvature scale, $\mu << L$, the width for decay into the bulk
dimension becomes
\begin{equation}
\Gamma_{\rm fermion}= \biggl({m_0}/{2L}\biggr)^{2gv/L}
({2 \pi L}/{[{\it \Gamma}(gv/L+1/2)]^2})~~.
\label{gammaf1}
\end{equation}
where, ${\it \Gamma}$ on the r.h.s. is the normal gamma function.
In the limit, $\mu >> L$ one similarly obtains
\begin{equation}
\Gamma_{\rm fermion} = M \biggl({m_0}/{2M}
\biggr)^{2M/L }
\exp{\{2 {M}/{L}\}}~~, 
\label{gammf2}
\end{equation}
where $M = \sqrt{(gv)^2 + \mu^2}$.

Clearly, in each of these expressions, the largest width for tunneling
into the bulk dimension is for the heaviest particle. In this case we
argue that a heavy ($^>_\sim$ TeV) dark matter particle [e.g. the
lightest supersymmetric  particle, (LSP)]  may have the shortest
lifetime to tunnel into the bulk. In this paper, therefore, we consider
the possibility that cold dark matter (CDM) disappears into the extra
dimension. The comoving density of the CDM will then diminish over time as
$(\rho_{\scriptscriptstyle CDM} a^3)\exp{[- \Gamma t ]}$.

 In principle, normal standard-model particles (e.g.~baryons) would
decay in this way as well. This would have many far reaching
consequences in astrophysics and cosmology.  However, the decay width
of such light particles is likely to be suppressed relative to that of a
heavy dark-matter particle by some power of the ratio of their masses
[e.g. by $(m_{baryon}/M_{LSP})^{2gv/L}  \sim (0.001)^{2gv/L}$ for a TeV
fermion (e.g. neutralino) LSP]. We also note, that even a  light
(axion-like) scalar dark matter particle could also be made to  have a
short disappearance time relative to normal fermionic matter (by
Eq.~\ref{gammaf1}) as long as  $(m_0/2 L) < 1$, and $gv/L$ is
sufficiently large to suppress the disappearance of normal fermionic matter.

In what follows, we analyze cosmological constraints on such
disappearing dark-matter particles and show that this hypothesis is
consistent with and even slightly preferred by all cosmological
constraints, including primordial nucleosynthesis, the present
observations of Type Ia supernovae at high redshift, the mass-to-light
ratio vs.~redshift relation of galaxy   clusters, the fraction of X-ray
emitting gas in rich galactic clusters and the cosmic microwave
background (CMB).

Cosmological constraints on decaying matter have been considered in many
papers, particularly with regard to the effects of such decays on
big-bang nucleosynthesis (cf. \cite{malaney,kawasaki} and
Refs. therein).  The present discussion differs from the previous
considerations in that the decaying particles do not produce photons,
hadronic showers, or residual annihilations in our four-dimensional
spacetime.  To distinguish  the disappearance of dark matter in the
present application from the previous decay applications, we shall refer
to it here as {\it disappearing dark matter}.

In the present application, however, there are some complications.
One is that, an energy flow into the bulk can induce a back reaction
from the background gravitational field. This leads to residual gravity
waves in the 3-brane from the exiting particles \cite{Gregory}.  Another
effect is an enhanced electric part of the bulk Weyl tensor \cite{Langlois}.
Together these effects will comprise the so-called ``dark radiation" as
analyzed below.  Another consideration is that particles which enter the
bulk can still interact gravitationally with particles on the brane.
The strength of this interactions, however, is greatly  diminished
\cite{Giddings} by a factor of $(R/z)$, where $z$ is the distance
between the bulk and the brane, and $R = 1/L$ is the "radius" of the
bulk dimension.  For a typical value of $L= 10^4$ GeV, we have $R \sim
10^{-4}$  fm.  So, even though gravity can reside in the bulk, the
residual gravity between particles in the bulk and brane is strongly
suppressed.

\section{Cosmological Model}
The five-dimensional Einstein equation for the brane world can be
reduced to an effective set of four-dimensional equations on the brane 
\cite{Shiromizu, Yokoyama, Tanaka} by decomposing the five-dimensional
Riemann tensor into a  Ricci tensor plus the five dimensional Weyl tensor.
The four-dimensional effective energy-momentum tensor contains the usual 
$T_{\mu\nu}$ term of ordinary  and dark matter plus a new term quadratic
in $T_{\mu\nu}$, and a residual term containing the five-dimensional Weyl
tensor with two of its indices projected along a direction normal to the
brane. The (0,0) component of the effective four-dimensional  Einstein
equation can then be reduced to a new generalized Friedmann equation 
\cite{Kraus,Binetruy,Hebecker,Ida,Mukohyama,Vollick}
for the Hubble expansion as detected by an observer on the three brane, 
\begin{equation}
H^2 = \left(\frac{\dot{a}}{a}\right)^2
=\frac{8 \pi G_{\rm N}}{3} (\rho + \rho_{\scriptscriptstyle DR})
-\frac{k}{a^2}+\frac{\Lambda_{4}}{3}
+\frac{\kappa_{5}^4}{36}\rho^2 ~~.
\label{Friedmann}
\end{equation}
Here,  $a(t)$ is the scale factor at cosmic time $t$, and
$\rho=\rho_B + \rho_\gamma + \rho_{\scriptscriptstyle DM}$, with 
$\rho_B$ and $\rho_\gamma$ the usual contributions from nonrelativistic
(mostly baryons) and relativistic particles, respectively.
In the present application we presume that only the dark matter can
decay into the extra dimension.  Hence, we write $\rho_{\scriptscriptstyle DM}=Ce^{-\Gamma
t}/a^3$, where $\Gamma$ is the decay width into the extra dimension.

In equation (\ref{Friedmann}), several identifications 
of cosmological parameters were required in order to recover standard
big-bang cosmology. For one, the first term on the right hand side
is obtained by relating the four-dimensional gravitational constant
$G_{\rm N}$ to the five-dimensional gravitational constant,
$\kappa_{5}$.  Specifically,
\begin{equation}
G_{\rm N} = M_4^{-2} = \kappa_{5}^4 \tau / 48 \pi~~,
\label{gn}\end{equation}
where $\tau$ is the brane tension and 
\begin{equation}
\kappa_5^2 = M_5^{-3}~~.
\label{kappa}
\end{equation}
where $M_5$ the five-dimensional Planck mass. Secondly, the
four-dimensional cosmological constant $\Lambda_{4}$ is related to its
five-dimensional counterpart $\Lambda_{5}$,
\begin{equation}
\Lambda_{4}=\kappa_{5}^4 \tau^2 /36 + \Lambda_{5}/6~~.
\label{lambda4}
\end{equation}
A  negative $\Lambda_{5}$  (and $\kappa_{5}^4 \tau^2 /36 \approx  \vert
\Lambda_{5}/6\vert$) is required for $\Lambda_{4}$ to obtain its
presently  observed small value.

Standard big-bang cosmology does not contain the $\rho_{\scriptscriptstyle DR}$ and $\rho^2$
terms of Eq. (\ref{Friedmann}). The $\rho^2$ term arises from the
imposition of a  junction condition for the scale factor on the surface
of the brane. Physically, it derives from the fact that matter fields
are initially confined to the brane. This  term decays rapidly as
$a^{-8}$ in the early radiation dominated universe and is not of
interest here.

In the present formulation, $\rho_{\scriptscriptstyle DR}$ includes two contributions,
$\rho_{\scriptscriptstyle DR} = \rho_{\scriptscriptstyle E} + \rho_{\scriptscriptstyle GW}$. One is the $\rho_{\scriptscriptstyle E}$ term which
derives from the electric part of the Bulk Weyl tensor. The second
($\rho_{\scriptscriptstyle GW})$ arises from residual gravity waves left on the brane
\cite{Gregory}. Since these gravity waves are associated with the
disappearing particles, their dynamics can be formally absorbed together
with $\rho_{\scriptscriptstyle E}$ into a Bianchi identity for the effective four-dimensional
Einstein equation. This leads to,
\begin{equation}
{\dot \rho_{\scriptscriptstyle DR}} + 4H \rho_{\scriptscriptstyle DR} = 
\Gamma \rho_{\scriptscriptstyle DM}~~.
\label{Bianchi}
\end{equation}
When $\Gamma=0$, $\rho_{\scriptscriptstyle DR}$ scales as $a^{-4}$ like normal radiation
even though it has nothing whatsoever to do with electromagnetic radiation.
Hence, the name `dark radiation'. Upper and lower limits on such dark
radiation can be deduced from big-bang nucleosynthesis \cite{Ichiki}.
In the present paper we will keep the same name, even though
in this more general context $\rho_{\scriptscriptstyle DR}$ no longer scales as $a^{-4}$. 

\begin{figure}
\rotatebox{0}{\includegraphics[width=0.35\textwidth]{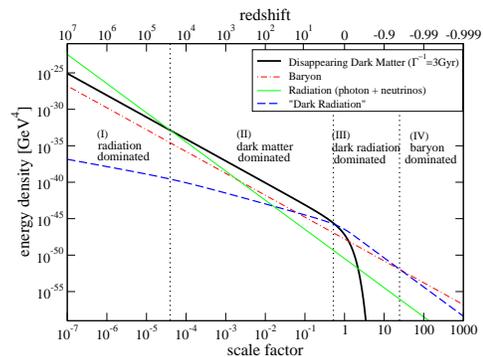}}
\caption{Illustration of the energy densities with scale factor
in models with dark-matter decay into the extra dimension.}
\label{fig:1}
\end{figure}

The introduction of the dark radiation term  into Eq. (\ref{Friedmann})
leads to new cosmological paradigms. For example, Figure \ref{fig:1}
illustrates the evolution of a a simple flat, $\Lambda_4 = k = 0$,
disappearing dark matter cosmology with negligible $\rho^2$ term. This
cosmology separates into four characteristic regimes identified on
Figure \ref{fig:1}. These are: I) The usual early radiation dominated
era ($z  > 10^5$); II) a dark-matter dominated era ($t << 2
\Gamma^{-1}$, $ 10 < z <10^5$); III) a late dark radiation dominated era
($t>> 2 \Gamma^{-1}$, $0 < z < 0.2$); and IV) Eventually, a
baryon-dominated regime also exists.

Early on the contribution from the  dark radiation component evolves
(from Eq.(\ref{Bianchi})) as $\rho_{\scriptscriptstyle DR} \propto a^{-1}$ 
or $\rho_{\scriptscriptstyle DR}
\propto a^{-3/2}$ during regimes I and II, respectively, and can be
neglected. Thus, the dark radiation does not affect (nor is it
constrained by) primordial nucleosynthesis. Similarly, the dark
radiation does not contribute  much mass energy during  the epoch of CMB
photon decoupling (at $z \sim 10^3$), though it  can become comparable
to and even in excess of the dark matter contribution in epoch III and
therefore affects the look-back time to the CMB epoch.

The most interesting region for our purpose is during the transition
from epoch II to epoch III. This occurs at intermediate times $t \sim
2\Gamma^{-1}$ and redshifts of $0 < z < 2$ as indicated on Figure
\ref{fig:1}. Here, the fact that there is both more dark matter  and
more dark energy at higher redshifts means that the universe decelerates
faster during the redshift regime $1 < z < 2$ than during the more
recent epoch $0 < z < 1$.   As far as cosmological
constraints are concerned, the most important effect is from the
changing dark matter contribution.  This is because  the dark radiation does not
become significant until the most recent ($z ^<_\sim  0.05$) epoch
even for this extreme cosmology.
The changing dark matter  contribution in particular,
 can  nevertheless have important observable consequences, for
example  on the
luminosity-redshift relation, galaxy mass-to-light ratios, and the
cosmic look-back time.  Hence, this model is constrainable by the
observations of supernovae and galaxy mass-to-light ratios at high
redshift, and the power spectrum of the  cosmic microwave background as
we now show.

\section{Supernova Constraint}
The apparent brightness of the Type Ia supernova standard candle with
redshift is given \cite{Carroll} by a simple relation which we slightly
modify to incorporate the brane-world cosmology given in
Eq.~(\ref{Friedmann}). The luminosity distance becomes,
\begin{eqnarray}
D_L &=& {c (1+z) \over H_0  \sqrt{\Omega_k}} sinn 
\biggl\{ \sqrt{\Omega_k} \int_0^z dz'[\Omega_\gamma (1 + z')^4 
\nonumber \\
&& 
+ (\Omega_{\scriptscriptstyle DM}(z')
 + \Omega_B) (1 + z')^3 
\nonumber \\
&& 
 + \Omega_k(1 + z')^2
 + \Omega_\Lambda+  \Omega_{\scriptscriptstyle DR}(z') \biggr]^{-1/2}  \biggr\}~~,
\end{eqnarray}
where $H_0$ is the present value of the Hubble constant, and $sinn(x) =
\sinh{x}$ for $\Omega_k > 0$, $sinn(x)=x$, for  $\Omega_k = 0$  and
$sinn(x) = \sin{x}$ for $\Omega_k < 0$. The $\Omega_i$ are the usual
closure quantities, i.e. the contribution from all relativistic
particles is $\Omega_\gamma = {8 \pi G \rho_\gamma / 3 H_0^2}$, the
baryonic contribution is $\Omega_B = {8 \pi G \rho_{B} / 3 H_0^2} =
0.039$ \cite{Ichiki} (for $H_0 = 71$ km s$^{-1}$ Mpc$^{-1}$).  The
curvature contribution is $\Omega_k = -{k / a_0^2 H_0^2}$, and
$\Omega_\Lambda  = {\Lambda / 3 H_0^2}$ is the vacuum energy
contribution. In the present context, we have added a redshift-dependent
contribution from the dark radiation, $\Omega_{\scriptscriptstyle DR}=  {8 \pi G
\rho_{\scriptscriptstyle DR}(z)/ 3 H_0^2}$.  The dark matter contribution $\Omega_{\scriptscriptstyle DM}$
becomes a function of redshift through, $\Omega_{\scriptscriptstyle DM} \longmapsto
\Omega_{\scriptscriptstyle DM}^0 \exp{\{\Gamma (t_0-t)\}}$, where $\Omega_{\scriptscriptstyle DM}^0 = {8 \pi G
\rho_{\scriptscriptstyle DM}^0 / 3 H_0^2}$ is the  present dark-matter content, and the
look-back time $t_0 - t$ is a function of redshift,
\begin{eqnarray}
t_0 -  t &=&  H_0^{-1}  \biggr\{ \int_0^z (1 + z')^{-1} 
 \biggl[\Omega_R (1 + z')^4
 \nonumber \\
& +& (\Omega_{B} +  \Omega_{\scriptscriptstyle DM}) (1 + z')^3 
\nonumber \\
 &+& \Omega_k(1 + z')^2
 + \Omega_\Lambda 
 + \Omega_{\scriptscriptstyle DR}
\biggr]^{-1/2} dz' \biggr\}~~.
\end{eqnarray}

\begin{figure}
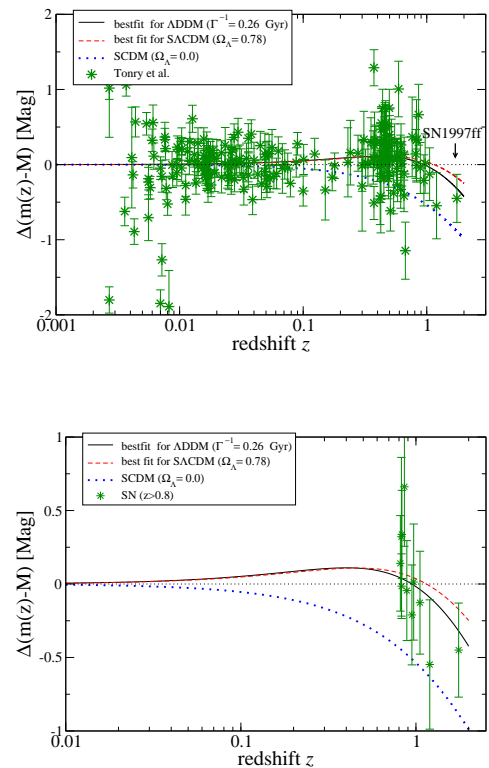

\rotatebox{0}{\includegraphics[width=0.35\textwidth]{mzwithdr.v11.eps}}

\vspace*{10mm}
\rotatebox{0}{\includegraphics[width=0.35\textwidth]{mzwithdr.zgt0.8.eps}}
\caption{Illustration of the supernova magnitude redshift
relation for various cosmological models with and without 
disappearing dark matter as labeled. 
The upper figure shows the full data set of \cite{tonry}.
The lower figure highlights the points with $z > 0.8$ most relevant to this paper.}
\label{figsn}
\end{figure}

Figure \ref{figsn} compares various cosmological models with some  of
the recent combined data from the High-Z Supernova Search Team
\cite{garnavich,tonry} and the Supernova Cosmology Project
\cite{perlmutter}. The lower figure highlights the crucial data points
at the highest redshift which are most relevant to this study. Shown are
the K-corrected magnitudes  $m = M + 5 \log{ D_L} + 25$
vs.~redshift. Curves are plotted relative to an open $\Omega_{\scriptscriptstyle DM},
\Omega_B , \Omega_{\Lambda}, \Omega_{\scriptscriptstyle DR} = 0$, $\Omega_k  = 1$
cosmology. Of particular interest are the highest redshift points
(e.g. SN1997ff \cite{Riess,tonry} at  $z = 1.7$). These points
constrain the redshift evolution during the important dark-matter
dominated decelerating phase relevant to this paper.

It is noteworthy that an optimum standard  flat $\Omega_M =
0.3$, $\Omega_\Lambda = 0.7$ (${S\Lambda CDM}$) cosmology passes
somewhat above the five points with $z\ge 0.9$. Indeed, the newest "Fall
1999" data \cite{tonry}  (shown in the lower box of Figure \ref{figsn})
are consistently brighter than the best-fit standard 
flat ${S\Lambda CDM}$ cosmology in the epoch at $z > 0.9$. 
This is made more relevant in view of the fact that dust around
SN1997ff would cause that inferred data point to be even lower on
this plot  \cite{Riess}. Thus, we find that the data
all slightly favor the disappearing dark matter ($\Lambda DCDM$)
cosmology.

\begin{figure}
\rotatebox{-90}{\includegraphics[width=0.35\textwidth]{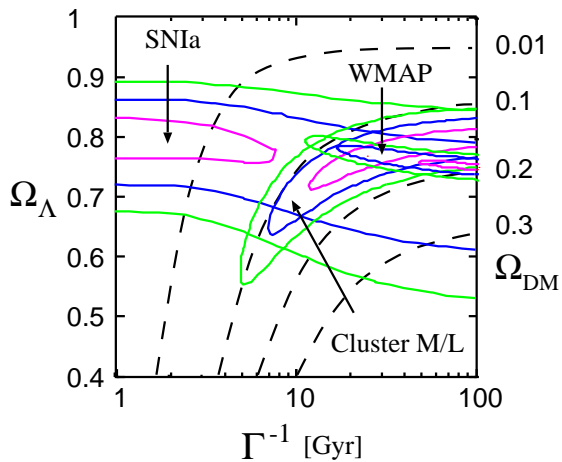}}
\caption{Contours of constant $\chi^2$ in the $\Gamma^{-1}$
 vs. $\Omega_\Lambda$ plane. Lines drawn correspond to 1, 2, and
 3$\sigma$ confidence limits for fits to the magnitude-redshift relation
 for Type Ia supernovae, the mass-to-light ratios of galaxy   clusters,
 and constraints from the CMB. The dashed lines indicate contours of
 constant $\Omega_{\scriptscriptstyle DM}$ as labeled. The dark radiation contribution can
 be deduced from the figure, via
 $\Omega_{\scriptscriptstyle DR}=1-\Omega_\Lambda-\Omega_{\scriptscriptstyle DM}-\Omega_B$.
}
\label{figcont}
\end{figure}

The contours labeled SNIa of Figure \ref{figcont} show 1$\sigma$,
2$\sigma$, and 3$\sigma$ confidence limit regions of constant goodness
of fit to the $z > 0.01$ data of \cite{tonry} in the parameter space
of disappearance lifetime $\Gamma^{-1}$ versus $\Omega_\Lambda$ plane.
For these data we use  a simple $\chi^2$ measure of the goodness of fit
as in \cite{tonry}.
\begin{equation}
\chi^2 \equiv \sum (Y_i^{data} - Y_i^{calc})^2/ \sigma_i^2~~,
\end{equation}
 where, $\sigma_i$ includes the velocity uncertainty added 
to the distance error.

The  SNIa data imply a shallow minimum  for $\Gamma^{-1} \approx 0.3  $
Gyr  and $\Omega_\Lambda = 0.78$. The reduced $\chi^2$ per degree of
freedom at this minimum is $\chi^2_r = 0.94$  for 171 degrees of freedom.
This is to be compared with compared with $\chi^2_r =
0.96$ for a standard $\Lambda CDM$ cosmology \cite{tonry}. 
The $1 \sigma $ confidence limit
corresponds to $\Gamma^{-1} \le 10 $ Gyr, but the $2 \sigma $ region is
consistent with a broad range of $\Gamma$ as long as $\Omega_{\Lambda} =
0.75 \pm 0.15$.

\section{Galaxy Cluster $M/L$  Constraint}
Another interesting cosmological probe comes from galaxy cluster
mass-to-light ratios as also shown on Figure \ref{figcont}.  This is the
traditional technique to obtain the total universal matter content
$\Omega_M$.  A most recent average value of $\Omega_M = 0.17 \pm 0.05$
has been determined in \cite{bahcall} based upon 21 galaxy   clusters
out to $z \approx 1$ corrected for their color  and evolution with redshift.
The very fact that the nearby cluster data seem to prefer a smaller value
of $\Omega_M$ than the value of $\Omega_M = 0.27 \pm 0.02$ deduced
\cite{WMAP} from the distant microwave background surface of photon last
scattering is consistent with the notion of disappearing dark matter as
discussed below.

\begin{figure}
\rotatebox{0}{\includegraphics[width=0.35\textwidth]{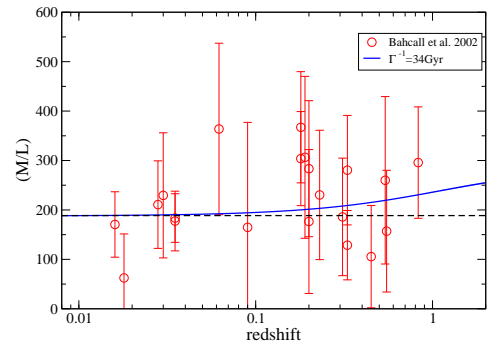}}
\caption{Illustration of the evolution and temperature corrected 
galaxy cluster mass-to-light ratios (from \cite{bahcall}) as a function of
 redshift.  The solid line shows the best fit cosmology with
 disappearing dark matter as described in the text.  The dashed line
 shows the present value of $\Omega_{M}$ as deduced from the nearby
 cluster data.}
\label{figclust}
\end{figure}

In the present disappearing dark matter paradigm, the dark matter
content diminishes with time, while the normal baryonic luminous matter
remains mostly confined to the brane. Therefore, the $M/L$ ratio should
increase with look-back time. This is complicated, however,  by two
effects.  One is that clusters at high redshift have had less time to
evolve and dim. Hence, their $M/L$ ratios are expected to decline with
redshift. This effect is corrected in Table 1 of \cite{bahcall}.
Another complication is an observational bias due to the fact that
at high redshift a larger fraction of high-temperature clusters is
observed.  In essence, higher temperature clusters have deeper
gravitational wells and are expected to have more dark matter and larger
$M/L$ ratios. Nevertheless, we have corrected for this temperature bias
by using the power-law analysis described in \cite{bahcall} to adjust
all clusters to a common temperature. Even after applying this correction 
we find a residual trend of increasing cluster $M/L$ ratio with redshift
which can be attributed to disappearing dark matter as depicted in
Figure \ref{figclust}.

Our standard $\chi^2$ goodness of fit to the data of \cite{bahcall}
(corrected for evolution and temperature bias) is labeled as Cluster M/L
on Figure \ref{figcont}. We find a minimum $\chi^2$ per degree of
freedom of $\chi^2_r = 0.61$ for 
$\Gamma^{-1} = 34$ Gyr as shown on Figures \ref{figcont} and \ref{figclust}. This is an
improvement over the fit with a fixed $M/L$ (shown as the straight
dashed line on Figure \ref{figclust}) for which $\chi^2_r = 0.67$. The
2$\sigma$ (95\% confidence level) limits from the galaxy cluster data
correspond to $\Gamma^{-1} \ge 7$ Gyr for our flat $\Lambda{DCDM}$ model
as shown in Figure \ref{figcont}. This limit is concordant with the
previously discussed Type Ia supernova analysis.

\begin{figure}
\vskip .1 in
\rotatebox{0}{\includegraphics[width=0.35\textwidth]{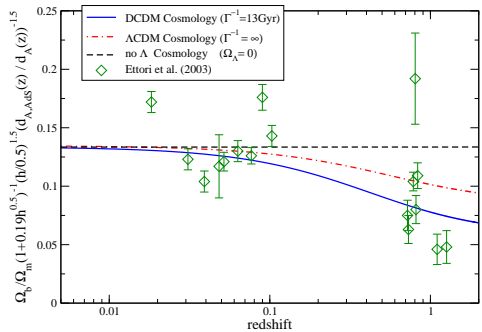}}
\caption{Illustration of the fraction of X-ray emitting gas to total
 mass  in rich clusters as a function of  redshift  (from \cite{ettori})
 as a function of redshift.  
 The solid line shows the theoretical gas mass fraction from the 
 disappearing dark matter cosmology as described in the text. The dashed line  
 shows a $\Lambda = 0$ cosmology and the dot-dashed line is for a    
 standard $\Lambda CDM$ cosmology. All theoretical models are        
 normalized to have the same gas mass fraction at present ($z=0$).  
}
\label{figxray}
\end{figure}

Clearly, more work is needed to unambiguously identify evidence for
enhanced dark matter in the past.  In this regard we note that there is
complementary data \cite{ettori} to the cluster $M/L$ ratios from  {\it
BeppoSax} and the {\it ROSAT} X-ray observations of rich clusters at
high redshift. In this case, the X-ray emitting gas mass can be
determined from the X-ray luminosity and the total mass deduced from the
gravitational mass required to maintain the X-ray gas in hydrostatic
equilibrium. There is, however, uncertainty in this method due to the
model dependence of the inferred gas fractions \cite{ettori}.
Nevertheless, the observations clearly exhibit a trend of diminishing
gas fraction for systems with $z > 1$. Figure \ref{figxray} shows a
comparison of the deduced gas fractions for various cosmological
models. These data are consistent with an increasing total mass
content for these systems as predicted in this disappearing dark matter
paradigm.

\section{CMB Constraint}
As noted above, the matter content ($\Omega_M = 0.27 \pm 0.02$) deduced
from the recent high-resolution {\it WMAP} analysis \cite{WMAP} of the
cosmic microwave background is larger than that deduced ($\Omega_M =
0.17 \pm 0.05$) from nearby galaxy cluster mass-to-light ratios
\cite{bahcall}.  This in itself is suggestive of the disappearing dark
matter paradigm proposed here.  However, this cosmology can also
involve a shorter look back time and different expansion history between
now and the epoch of photon last scattering. In particular there will be
more dark matter at earlier times leading to earlier structure
formation. 
There will also be a smaller integrated Sachs-Wolf effect (ISW) at early times, 
and a larger ISW effect at late times as photons propagate to the present epoch.   
Thus, the amplitudes and locations of the peaks in the power spectrum
of microwave background fluctuations \cite{hu} can in principle be used
to constrain this cosmology.

We caution, however, that there is a complication with using the CMB
constraint.  Inflation generated metric fluctuations which contribute to
the CMB should also induce fluctuations in the dark radiation component.
Unfortunately, however, calculations of the power spectrum from five
dimensional gravity are complicated and beyond the scope of the present
work. A straight forward application of this disappearing dark
matter paradigm without a proper treatment of the fluctuation power spectrum
from the dark radiation should  therefore probably be viewed with caution.
Nevertheless, under the assumption that
fluctuations in the dark radiation contribute
insignificantly to the power spectrum at the surface of photon last
scattering, a straight forward study of the CMB  constraints on the
disappearing dark matter cosmology is possible.

\begin{figure}
\rotatebox{0}{\includegraphics[width=0.35\textwidth]{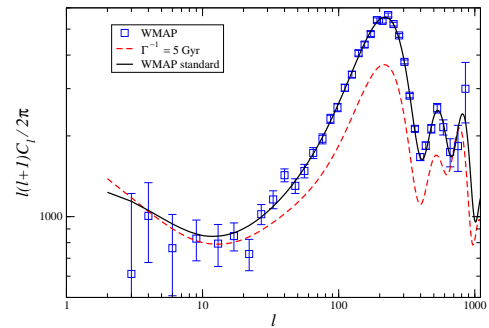}}
\caption{Illustration of a disappearing dark matter cosmology (dashed
 line) which is ruled out by the {\it WMAP} power spectrum. The points
 are the {\it WMAP} data.  The solid line is the standard best fit
 \cite{WMAP} for a normal $\Lambda CDM$ cosmology.
}
\label{figspec}
\end{figure}

We have done calculations of the CMB power spectrum, $\Delta T^2 =
l(l+1)C_l/2 \pi$ based upon the CMBFAST code of Seljak \& Zaldarriaga
\cite{cmbfast}. We have explicitly modified this code to account for the
the disappearing dark matter cosmology described in Eq. (\ref{Friedmann}).
Figure \ref{figspec} shows an illustration of a disappearing dark matter
model which can be ruled out by the CMB. In this example $\Gamma^{-1} =
5$ Gyr, and all other cosmological parameters set to their best fit {\it
WMAP} values \cite{WMAP}.

Nevertheless, it is quite possible to have a finite $\Gamma$ and still
fit the {\it WMAP} data. 
As an illustration of this, we have simultaneously varied     
$\Gamma$ and $\Omega_\Lambda$, and marginalized over the parameters   
of the matter power spectrum, while maintaining other cosmological    
parameters at the best fit {\it WMAP} values.                         
The likelihood functions we computed from a combination of Gaussian 
and lognormal distributions as described in Verde et al.~\cite{Verde}.                                                         
These were used to generate contours of  1, 2, and 3$\sigma$ confidence
limits for fits to the {\it WMAP} power spectrum \cite{Hinshaw,Kogut}
as shown on Figure \ref{figcont}.  

An important point is that we find that equivalent fits to that of the
best-fit {\it WMAP} parameters \cite{WMAP} can be obtained  for a broad
range of values for $\Gamma$ and $\Omega_\Lambda$. This means that the
CMB does not rule out this paradigm. On the contrary, the 2$\sigma$ CMB
contours nicely overlap the region allowed by the cluster $M/L$ ratios.
A $2\sigma$ concordance  region of $ 15 \le \Gamma^{-1} \le 80$ Gyr
survives this constraint. The essential requirements to fit the CMB in
this model is that the matter content during photon decoupling be at the
(higher) {\it WMAP} value, and that the dark radiation be an
insignificant contributor to the background energy density  during that
epoch.
\section{Conclusion}
Obviously,  there is great need for better Type Ia supernova data in the
crucial $z > 1$ regime as well as more galactic cluster mass-to-light
ratios at high redshifts. Although the evidence for disappearing dark
matter is of marginal statistical significance at the present time, the
purpose of this paper is nevertheless to emphasize the potential
importance of future studies aimed at unambiguously determining the
decay width. If such a finite value of $\Gamma$  were to be established,
it would constitute the first observational indication for noncompact
extra dimensions.  It would also provide valuable insight into the
physical parameters of the higher-dimensional space. 

Rewriting the equation for the decay width, along with the relations
(Eqs. \ref{gn} - \ref{lambda4}) between various quantities in the
modified Friedmann equation, i.e. $\kappa_5$, $G_N$, $M_4$, $M_5$,
$\Lambda_4$, and $\Lambda_5$, leads to the following relation between
the five-dimensional Planck mass $M_5$ and quantities which can be
measured in the four dimensional space time, $M_5^6 =
({M_4^4}/{64\pi^2})[{\pi}m_0^3/{16\Gamma}+\Lambda_4]$. Other fundamental
parameters in five dimensions, e.g.~$\Lambda_5$ and the brane tension
$\tau$, are derivable from $M_5$ via equations (\ref{gn}) and
(\ref{lambda4}). This implies that, should the dark-matter mass $m_0$
ever be known, all of the five-dimensional parameters could be
determined. For example, a dark matter mass of $m_0 \approx  1$ TeV (as
expected for the LSP), and a most optimistic decay lifetime of
$\Gamma^{-1} = 15$ Gyr, would imply $({M_5}/{M_4}) \approx 4 ({m_0}/
{\rm TeV})^{1/2} ({\Gamma^{-1}}/{15~{\rm Gyr}})^{1/6}$.

\acknowledgments
One of the authors (MY) would like to thank K. Ghoroku for helpful
discussions. We also acknowledge Y. Fujita, T. Tanaka and Y. Himemoto for
discussions which improved an earlier version of this paper. Work
at NAOJ has been supported in part by 
the Sasakawa Scientific Research Grant from the Japan Science Society,
and also by the
Ministry of Education, Science, Sports and Culture of Japan
through Grants-in-Aid for Scientific Research (12047233, 13640313,
14540271), and for Specially Promoted Research (13002001).
 Work at the University of Notre Dame was supported by the 
U.S.~Department of Energy
under Nuclear Theory Grant DE-FG02-95-ER40934.

\end{document}